\documentclass[twocolumn,aps]{revtex4-1}

\usepackage{graphicx}
\usepackage{amssymb}
\usepackage{amsfonts}
\usepackage{amsmath}
\usepackage{amsbsy}
\usepackage{natbib}
\usepackage{comment}
\usepackage{color}
\usepackage[dvipsnames]{xcolor}

\usepackage{epstopdf}
\renewcommand{\vec}[1]{\mbox{\boldmath$\mathrm{#1}$}}
\def\ind#1{{_{\mathrm{#1}}}}
\newcommand{\be}{\begin{equation}}
\newcommand{\ee}{\end{equation}}
\newcommand{\ben}{\begin{eqnarray}}
\newcommand{\een}{\end{eqnarray}}

\begin{document}

\title{Size-dependent frequency bands in the ferromagnetic resonance of a  Fe-nanocube}
\author{A.F. Sch\"{a}ffer$^{1}$}

\author{A. Sukhov$^{1}$}
\email{alexander.sukhov@physik.uni-halle.de}

\author{J. Berakdar$^{1}$}

\affiliation{
$^{1}$Institut f\"ur Physik, Martin-Luther-Universit\"at Halle-Wittenberg, 06099 Halle (Saale), Germany 
}

\begin{abstract}
Using full micromagnetic simulations we calculate the spectra of ferromagnetic resonance (FMR) for an iron (core-shell) nanocube and show that the FMR characteristics are strongly size dependent. For instance, for a $40$~nm  it is found that, in contrast to a macrospin picture,  the spectrum of the iron nanocube possesses two bands  centered around $0.4$~T and $\approx 0.1$~T. The peaks originate from the surface anisotropy induced by the strong demagnetizing fields (DMFs) of iron. Further simulations reveal that for $\approx 20$~nm  nanocubes  the macrospin model becomes viable.  Above $40$~nm we find a broad band for FMR absorption. Our results point to possible interpretations of  existing FMR experimental observations for the system studied here.
\end{abstract}

\date{\today}

\maketitle

\section{Introduction}

 With the trend to miniaturization of magnetic structures, chemically prepared spherical nanoparticles of hard magnetic alloys like FePt have attracted intense attention in the past\cite{WeMo99,SuMu00}. Because of their strong magnetocrystalline anisotropy, they are potentially interesting for information storage in a single nanoparticle with a size  below $10$~nm, overcoming thus the superparamagnetic size limit at room temperatures.
  On the other hand, while their size distribution may  well be controllable, the orientation of the magneto-crystallographic directions for two-dimensional arrays of such spherical nanoparticles pose technical difficulties\cite{AnLi05}. To circumvent this problem, new wet-chemical methods were invented allowing to fabricate Fe/Fe$_\mathrm{x}$O$_{\mathrm{y}}$-core-shell-magnetic nanoparticles in form of cubes with side lengths ranging from $14$~nm\cite{TrMe08} or $18$~nm\cite{KrFr11} to over $40$~nm\cite{Terw12}. Forming nano cubes arrays  deposited onto a substrate sufficiently reduces the parameter space for characterization. Furthermore, as possessing weak magnetocrystalline anisotropy and having one of the largest magnetic moments among conventional ferromagnetic materials, it is expected that the shape and the surface effects will dominate the magnetization dynamics in Fe/Fe$_\mathrm{x}$O$_{\mathrm{y}}$-core-shell-nanocubes.

Some research works \cite{RaBa13} were already devoted to the investigation of emergent effects for nanoarrays in the FMR, however, not for nanocubes. In our previous studies\cite{SuHo14,HoSu15} we already performed detailed theoretical analysis of the influence of dipole-dipole interactions (DDI) on the spectra of ferromagnetic resonance (FMR)\cite{FaSi13} in arrays of ordered iron nanocubes with edge sizes of $40$~nm. However, this was carried out in the \textit{macrospin} approximation, ascribing to each nanocube a huge magnetic moment and neglecting the internal  magnetic structure of the nanocubes.
 Here, employing full micromagnetic simulation methods, we aim at examining the general  accuracy  of the macrospin model, and particularly how FMR spectra for a single Fe-nanocube will be modified.   We find that the macrospin approximation which was successfully employed for explanation of multiple peaks earlier \cite{PoRa16} generally is not suitable for the case discussed here.

\section{Model}

We consider a single iron-nanocube with a side length $a=40$~nm containing $N=1000$ cubic simulation cells, each having a size of $c=4$~nm. For cell sizes lower than the exchange length of bulk iron $l_{\mathrm{ex}}=\sqrt{A/(\mu_0 M_{\mathrm{S}}^2)}=2.4$~nm ($A$ is the exchange stiffness constant, $M_{\mathrm{S}}$ is the saturation magnetization), we did not observe any significant changes in the FMR-behavior. Each simulation cell is associated with a magnetization $\vec{m}_i=\vec{M}_i/M_{\mathrm{S}}$ normalized to $M_{\mathrm{S}}$. The dynamics of the magnetization $\vec{m}_i$ is governed by the Landau-Lifshitz-Gilbert equation \cite{LaLi35,Gilb55} $ d\vec{m}_i/dt =  - \gamma/(1+\alpha^2) [\vec{m}_i\times \vec{B}_i^{\mathrm{eff}}(t)] - \alpha \gamma/(1+\alpha^2) [\vec{m}_i\times [\vec{m}_i\times \vec{B}_i^{\mathrm{eff}}(t)]]$, where $\gamma=1.76 \cdot 10^{11}$~1/(Ts) is the gyromagnetic ratio and $\alpha$ stands for  the Gilbert damping. The local effective field is defined via $\vec{B}_i^{\mathrm{eff}}(t)=-1/M\ind{S}\delta F/(\delta \vec{m}_i)$, which is a function of the total free energy of the system $F = F\ind{EXCH} + F\ind{MCA} + F\ind{DMF} + F\ind{ZMN}$ (for details of each term we refer to Ref. \cite{SuHo14}). Further, $F\ind{EXCH}= - A/c^2\sum_{<ij>}\vec{m}_i \cdot \vec{m}_j$ describes the exchange interaction between nearest magnetic moments only, $F\ind{MCA}$ denotes the cubic magnetocrystalline anisotropy of bulk iron with the anisotropy axes' directions parallel to the cartesian ones. $F\ind{DMF}$ stands for demagnetizing fields and $F\ind{ZMN}=-\mu_0M\ind{S} \sum_i \vec{m}_i \cdot \vec{H}_{\Sigma}(t)$ is the Zeeman-energy ($\mu_0$ is the magnetic permeability of vacuum), where the total external magnetic field $\vec{H}\ind{\Sigma}(t)$ is a sum of a strong static $\vec{H}_{||}=H_{||}\vec{e}\ind{z}$ and a weak oscillating magnetic field $\vec{H}(t)=H_0 \cos \omega t \vec{e}\ind{x}$ with $\omega/(2\pi)=9$~GHz. Full GPU-based micromagnetic simulations using the simulation package \texttt{mumax3}\cite{VaLe14} were employed to account for the effect of demagnetizing fields.

The intensity detected in FMR-experiments\cite{Farl98} is related to the absorbed power when an oscillating field $\vec{H}(t)$ is applied.
 Defining the absorbed magnetic power density via $P = - \frac{\mu_0 M\ind{S}}{N} \sum_i \frac{1}{N\ind{T}T} \int_0^{N\ind{T}T} \vec{m}_i(t)\cdot \frac{\partial \vec{H}\ind{\Sigma}(t)}{\partial t}dt$\cite{Usad06,SuUs08}, where $N\ind{T}$ expresses the number of periods of calculations with respect to the external oscillating field, it can be shown\cite{OsEl14} that in the linear response regime $P $ is proportional to the imaginary part of the transverse magnetic susceptibility.

We employ material parameters related to  bulk iron \cite{Coey10} with the exchange stiffness $A=21 \cdot 10^{-12}$~J/m, the cubic anisotropy constants $K_{\mathrm{c} 1}=4.8 \cdot 10^{4}$~J/m$^{3}$, and $K_{\mathrm{c} 2}=-1.0 \cdot 10^{4}$~J/m$^{3}$, the saturation magnetization $M_{\mathrm{S}}=1.76 \cdot 10^6$~A/m and the damping constant $\alpha\approx~0.05$\cite{TrMe08}. To calculate FMR numerically, the simulations are performed at zero Kelvin by allowing the system to reach an equilibrium state after disregarding the first $90$ periods defined by the frequency of the external time-dependent magnetic field $\omega$.

\section{Numerical results and discussion}

\begin{figure}[!t]
\centering
\includegraphics[width=0.47\textwidth]{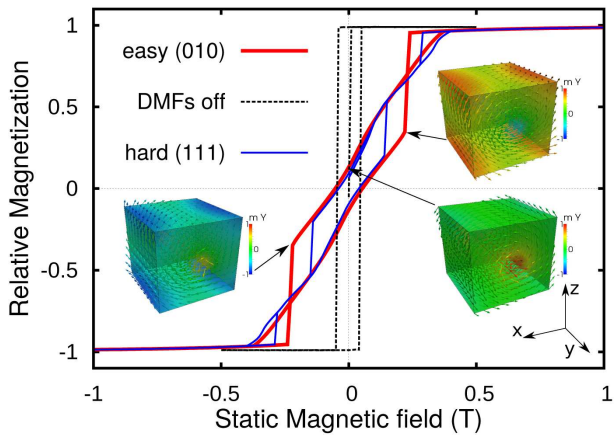}
\caption{Hysteresis curves for a single $40 \times 40 \times 40$ nm$^3$ Fe-nanocube plotted accounting (thick red and blue curves) or not (black curve) for the effect of the demagnetizing fields. Static magnetic field is applied either along the y-direction (easy axis) or along the main diagonal of the nanocube (hard axis). The oscillating field $\vec{H}(t)$ is not applied.}
\label{fig_1}
\end{figure}

\begin{figure}[!t]
\centering
\includegraphics[width=0.49\textwidth]{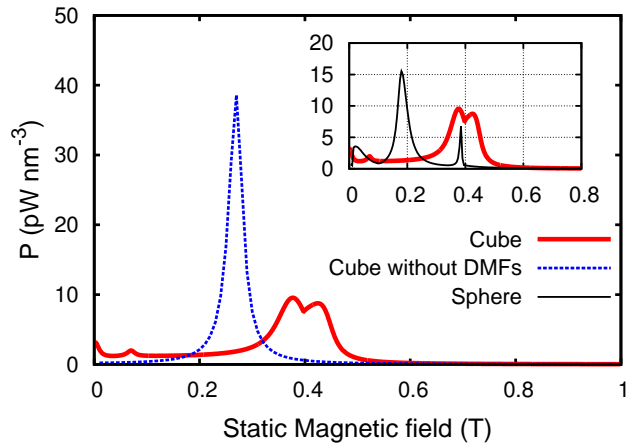}
\caption{Spectrum of absorbed power density for a single $40 \times 40 \times 40$ nm$^3$ Fe-nanocube. Inset shows a comparison with a spherical Fe-nanoparticle having a diameter of $40$~nm including the effect of demagnetizing fields.}
\label{fig_2}
\end{figure}

\begin{figure}[!t]
\centering
\includegraphics[width=0.49\textwidth]{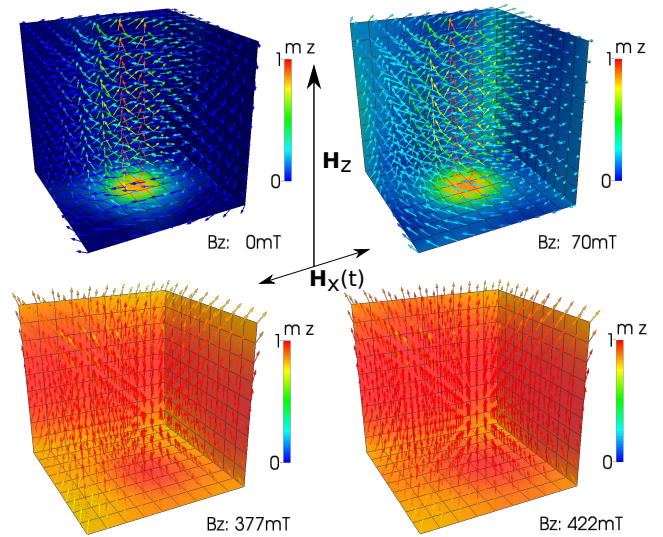}
\caption{Configurations of the initial state and the states corresponding to the three peaks of the cube's FMR-spectrum (cf. fig.\ref{fig_2}) and the schematics of the applied magnetic fields.}
\label{fig_3}
\end{figure}

While performing and analyzing the numerical calculations  effects caused by different initial magnetization states need to be considered carefully.
 As the system's free energy is minimized in the presence of the demagnetizing field, but without any applied external magnetic fields, the magnetization tends to a vortex like state (Fig. \ref{fig_1}). The associated orientation is randomly chosen out of the easy magnetization axes. For this reason an initial state dependence is present in the hysteresis loops, i.e. the beginning of the loop, but also in the FMR-spectra, where we can find minor peaks and leaps in the regime of field strengths about a few mT, if the initial curl is not oriented parallel to the static magnetic field. Hence, the direction of the vortex is attached very weakly.

First indications of noncollinear magnetization dynamics appear as the magnetic hysteresis is simulated. In this case a static magnetic field is applied either along the y-direction (easy magnetocrystalline axis) or along the main diagonal of the nanocube (hard magnetocrystalline axis). The resulting curl-like magnetization configuration possesses low magnetic remanence (Fig. \ref{fig_1}, red and blue curves). A further increase of the magnetic field leads to a growth of the net magnetization almost linearly, although after the saturation of the magnetization a jump in the net magnetization emerges, which is related to the switching of the corner and edge magnetic moments back to the direction of the field. The jumps of the net magnetization (spin configurations in Fig. \ref{fig_1}) are general for both the easy and the hard directions, however, quantitatively they differ in the value of the magnetocrystalline anisotropy field, the maximum of which can be estimated as $H^{\mathrm{max}}_{\mathrm{c}}=2K\ind{c1}/M\ind{S}\approx 0.056$~T. Driven by magnetocrystalline anisotropy only (DMFs=0) the obtained coercive field appears clearly close to the value $H^{\mathrm{max}}_{\mathrm{c}}$ (Fig. \ref{fig_1}, black, dashed curve). \\

When decreasing the external magnetic field starting from the saturated state, the net magnetization becomes negative ($m \approx -0.25$) in an almost linear slope, even before the system's configuration is switched. This aspect emerges due to the combination of the present magnetic stability of the center's magnetic moments and the more loosely arranged moments, shaping the twisting part. At first, only the curly part reacts on the applied field so that, directly before the switching point is reached, the configurations show a huge difference between the center's and the encircling moments' orientations (spin configurations in Fig. \ref{fig_1}).

Next, an FMR spectrum for a single Fe-nanocube under the conditions described in Sec. II is inspected (Fig. \ref{fig_2}, solid thick curve). We observe two main peaks of absorption around $0.4$~T and a small one below $0.1$~T (cf. corresponding states in Fig. \ref{fig_3}). In contrast to the macrospin case (shown in Fig. \ref{fig_2}, thin solid curve), whose peak's position is in line with the analytical estimate $\mu_0 H\ind{res} = \omega/\gamma - 2K\ind{c1}/M\ind{S} \approx 0.27$~T, the main peaks are shifted towards higher magnetic fields indicating involvement of an anisotropy. Clearly, this shift can not be caused by the magnetocrystalline anisotropy, since its maximum contribution of around $0.056$~T is too low. A more plausible candidate for the spectrum modification is the surface anisotropy of the cube's planes induced by the demagnetizing fields that force the magnetization to lie in-plane for all the surfaces. This guess is supported by a comparison with the spectrum of a sphere with diameter $40$~nm (inset of Fig. \ref{fig_2}). Generally, since a sphere has higher symmetry than a cube, resulting in zero net demagnetizing energy contribution \cite{Ohan00} $1/4\mu_0M^2_{\mathrm{S}}(1-3\frac{1}{3})=0$, the FMR for an ideal sphere yields a macrospin-like spectrum. Modeling of finite-size spheres with cubic cells, however, makes them flattened. As a result, we obtain the main peak at $0.2$~T, which is close to the macrospin's peak position, accompanied by two minor peaks at low and high fields presumably stemming from the flattened surfaces of the modeled sphere.

\begin{figure}[htb]
\centering
\includegraphics[width=0.48\textwidth]{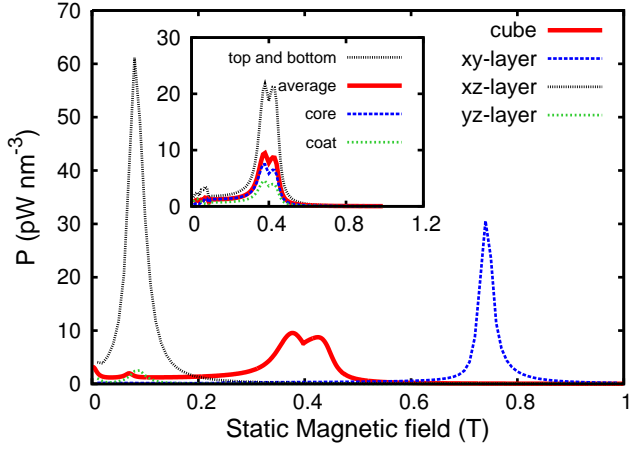}
\caption{Spectrum of the absorbed power density for a single $40 \times 40 \times 40$ nm$^3$ Fe-nanocube compared with the FMR-spectra of the  (xy-, yz-, xz-) cartesian planes calculated independently of being a part of the nanocube. Inset shows the contributions of the main planes being part of the nanocube.}
\label{fig_4}
\end{figure}

To unveil the origin of the double peak (Fig. \ref{fig_2}) we analyze the dynamics in the cartesian planes: the xy-,\\ xz- and yz-planes. In the  xz-plane, for instance (Fig. \ref{fig_4}, thin black dotted curve), we consider a layer having dimensions $ 40 \times 4 \times 40$~nm$^3$ and start from a relaxed state at zero magnetic field. The maximum associated with the xz-plane becomes shifted to the left with respect to the double peak. A similar behavior can also be observed for the yz-layer (Fig. \ref{fig_4}, green solid curve; size $4\times 40 \times 40$~nm$^3$) with a very low amplitude. The reason of having that different amplitudes for the xz- and yz-layers is due to a suppression of the x-component of the magnetization for the case of the yz-layer by the demagnetizing fields forcing it to be in the yz-plane, which acts against the driving field $\vec{H}(t)$ oscillating along the x-axis. Finally, the xy-layer yields a peak with a moderate amplitude shifted towards strong magnetic fields (Fig. \ref{fig_4}, blue dashed curve). Qualitatively, the positions of the emerging peaks of the respective planes can be elucidated by considering demagnetizing fields as a sort of surface anisotropy. Indeed, for the infinitely large xz- or yz-layers the surface anisotropy energy can have the form $+\mu_0M^2_{\mathrm{S}}/4m_{\mathrm{y}}^2$ and $-\mu_0M^2_{\mathrm{S}}/4m_{\mathrm{x}}^2$, respectively. For the infinitely long xy-layer we similarly arrive at $\mu_0M^2_{\mathrm{S}}/4m_{\mathrm{z}}^2$. Employing the definition for the effective field and the resonance condition $\mu_0H^{\mathrm{eff}}_{\mathrm{z}} = -\delta F /\delta m\ind{z}= \omega/\gamma$, one can easily achieve the resonance condition accounting among other contributions for the surface anisotropy induced by the demagnetizing fields $\mu_0 H^{\mathrm{res}} = \omega/\gamma - 2K_{\mathrm{c1}}/M\ind{S} + p_i \mu_0M\ind{S}/2$. Here, the factors $p_i=p_{\{\mathrm{xz-, yz-}\}}<0$ and $p_i=p_{\{\mathrm{xy-}\}}>0$ indicate  the peak positions for the corresponding planes and take their finite sizes into account.  
Similar conclusions can be achieved by considering the Kittel formula for the resonance frequencies (eq. (9) of Ref. \cite{Kitt48}). Indeed, for an infinitely large thin film with the external static field applied perpendicularly to the film, the position of the resonance is determined by $\mu_0 H_0 = \omega/\gamma + \mu_0 M$, which is consistent with our $xy$-case (cf. Fig. \ref{fig_4}). Analogously, we can qualitatively explain the positions of the resonance fields  in case of the $xz$- and $yz$-planes by assuming the Kittel formula in the situation of the static field being applied parallel to the infinitely large thin film plane, i.e. $(\omega/(\mu_0 \gamma))^2=H_0 (H_0 + M)$.

\begin{figure}[htb]
\centering
\includegraphics[width=0.49\textwidth]{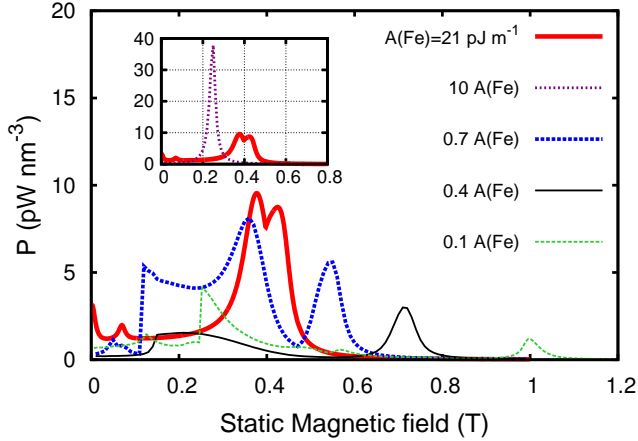}
\caption{Spectrum of the absorbed power density for a $40 \times 40 \times 40$ nm$^3$ Fe-nanocube calculated for different strengths of the exchange constant $A$.}
\label{fig_5}
\end{figure}

Additionally, selective contributions of the nanocubes' core, external layers or its top and bottom layers, when they are part of the nanocube, reveal no significant information (inset of Fig. \ref{fig_4}) to the verification of the FMR peaks, which leads to the conclusion that there should be a dependence of the peaks' positions on the exchange coupling constant. The latter issue is inspected in details in Fig. \ref{fig_5}. It demonstrates that when the exchange coupling is only ten percent of the bulk value of iron, it has one peak for approximately $1$~T and an increased intensity at low fields which has a leap around $0.2$~T. Such leap is caused by a rapid transition of the nanocube's magnetization curl along the x- or y-direction to the z-direction. Upon a further increase of the exchange constant the peaks gain in intensity and approach  the positions of those which are expected   for the nanocube's bulk value. The case of unrealistically strong exchange coupling (inset of Fig. \ref{fig_5}) illustrates the expected macrospin-like behavior.

\begin{figure}[htb]
\centering
\includegraphics[width=0.49\textwidth]{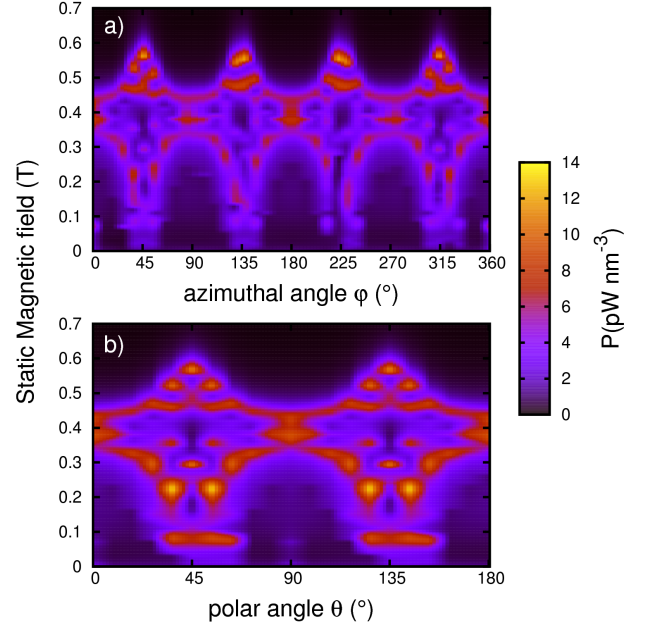}
\caption{Angular resolved FMR for a $40 \times 40 \times 40$ nm$^3$ Fe-nanocube. a) shows a $\varphi$-variation with fixed $\theta=\pi/2$. b) demonstrates a $\theta$-variation for fixed $\varphi=0$.}
\label{fig_6}
\end{figure}

The angular-resolved FMR performed for both polar and azimuthal angles (Fig. \ref{fig_6}) completes the picture of peaks evolution depending on the direction of the applied static field. In fact, it shows an envelope which is due to the magnetocrystalline anisotropy. Therefore, it has four-fold symmetry. Along the directions of the hard axes the number of peaks can even be higher than two, since the energy landscape is symmetric around the nanocube's edges and corners.

We notice in addition   to this envelope  also the multiple peaks pattern especially in the low field regime.
Hence the broad experimental spectra might also result from relatively small deviations of the cubes' orientations. Particularly when the magnetic field is pointing at the cube's edge the resonance field is sensitive to small angular variations (Fig. \ref{fig_6}).

\begin{figure}[!h]
\begin{center}
	\includegraphics[width=0.5\textwidth]{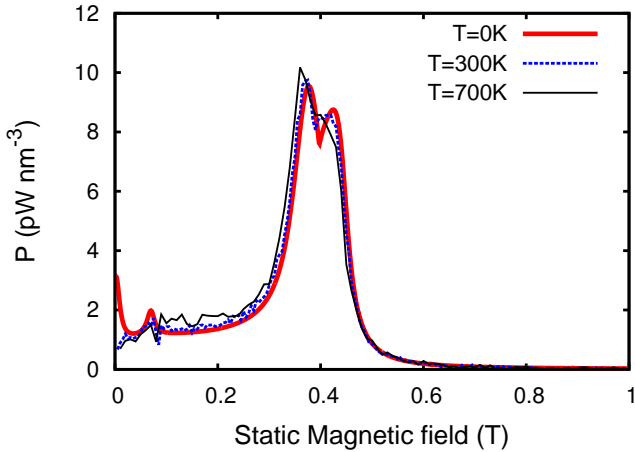}
	\caption{The effect of finite temperature on the FMR-spectra. The spectra are shown as a result of averaging over 200 cycles of the external AC field, i.e. $N_T=200$.}
	\label{fig_6a}
\end{center}
\end{figure}

So far only zero Kelvin calculations were discussed. When the local effective field is augmented by a stochastic field  $\vec{B}^{\mathrm{therm}}(t)$ with zero mean and the correlator $\left< B_{i\eta}(0)B_{j\theta}(t)\right> \propto \delta_{ij} \delta_{\eta  \theta}\delta(t) 2 \alpha k_B T/(\mu_S \gamma)$ ( $i$, $j$ stand for various simulation cells or magnetizations $\vec{m}_i$ and $\vec{m}_j$, $\eta$ and $\theta$ describe the cartesian components of the field, $\delta_{ij; \eta \theta}$ is the Cronecker notation and $\delta(t)$ represents the delta function; for details we refer to Ref. \cite{SuUs08}), then an averaging over multiple periods $N_{\mathrm{T}}=200$ of the oscillating magnetic field is indispensable. We point to a weak trend to shift the positions of resonances down to lower fields (Fig. \ref{fig_6a}) which can be referred to as an effective weakening of all deterministic interactions due to the presence of noise, i.e. the positions move towards the no DMFs-case. Nevertheless, the other features of the spectra remain the same, meaning that the double-peak splitting around $0.4$~T just as the width of those peaks, and the large background of the curve are preserved. The only noticeable  difference occurs for the regime of low field strengths ($B\leq 0.1$~T), where the thermal fluctuations suppress the minor peaks, created due to weak couplings or unstable configurations.

\begin{figure}[htb]
\centering
\includegraphics[width=0.49\textwidth]{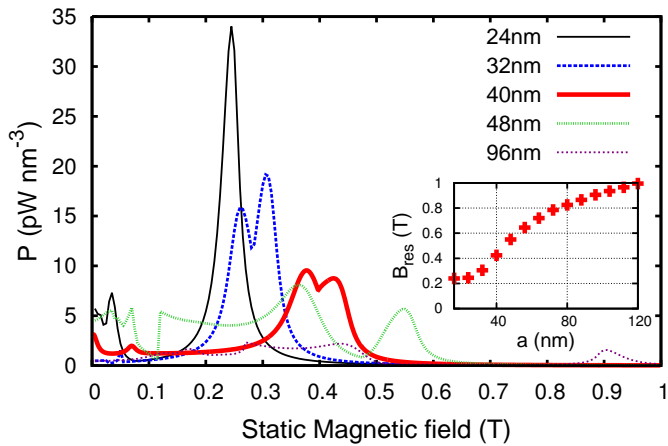}
\caption{Spectra for the absorbed power density for various sizes of the nanocube ranging from $20$ to $96$ nm of the cube's edge. Inset: Highest magnetic field corresponding to a resonance  is shown  as a function of the size.}
\label{fig_7}
\end{figure}

Our final point on interest concerns  finite-size effects (Fig. \ref{fig_7}). Apparently, the nanocubes with side sizes around $20$~nm behave similar to the macrospin scenario, since demagnetizing fields on the surfaces can not sufficiently develop and the six surfaces compensate each other identically with the case of a sphere. Noticeable in this respect is the fact that the appearance of multiple FMR-peaks is not necessarily connected with the  non-collinear configuration defined by the critical value of approximately  $8 l_{\mathrm{ex}}$ \cite{RaFa98} ($l_{\mathrm{ex}}^{\mathrm{Fe}}\approx 2.4$~nm). As explained earlier, the situation changes with a growing side size (over $40$~nm), indicating  an increasingly important  role of the demagnetizing fields that give rise to a significant surface anisotropy aligning the surface spins in-plane. Finally, for side sizes around $100$~nm the nanocube shows a broad range of absorption at low fields with a clear low-intensity peak at strong fields. This behavior is a signature of a complicated internal magnetization configuration.	 With increasing side sizes the inner vortex-like magnetization configuration exhibits a stronger protection against external fields, at the same time it does not show an outstanding collective resonance behavior any more. The larger vortex-structure screens thus the cube from the applied field.

\section{Conclusions}
We performed  extensive micromagnetic simulations for the magnetization dynamics of single $40\times 40 \times 40$~nm$^3$ Fe-nanocubes that were experimentally investigated and shown to form two-dimensional arrays of hundred nanocubes \cite{Terw12}. Accounting for local demagnetizing fields and temperature effects we analyzed the hysteresis behavior and the spectra of the absorbed power density in a ferromagnetic resonance setup. We observed a low remanence magnetization (Fig. \ref{fig_1}) explainable by the formation of a vortex magnetic state at zero applied field. This fact pointed to a strong impact of  the demagnetizing fields that resulted in two main peaks of FMR absorption located around $0.4$~T and a low-intensity peak at about $0.1$~T (Fig. \ref{fig_2}). As shown by Figs. \ref{fig_4}, \ref{fig_5} the peaks are traced back to the demagnetizing fields, the action of which can be interpreted as a sizable surface anisotropy. Surfaces with orientations along the xz- and yz-planes give rise to low-field peaks ($0.1$~T), whereas the xy-planes contribute to the absorption peaks at strong magnetic fields (about $0.7$~T, cf. Fig. \ref{fig_4}). Being coupled via exchange interaction in the nanocube the contributions of the planes merge to a doubled peak centered around $0.4$~T. In addition, as shown in the simulation for finite-size effects (Fig. \ref{fig_7}), generally, the macrospin approximation is good for side sizes around $20$~nm. Between $20$ and $40$~nm the nanocubes start developing numerous peaks of absorption, whereas above $40$~nm a complex FMR spectrum appears. For even larger cubes, further complications at  moderate field appear, but also a general shift of the maximum resonance peak can be seen (cf. inset of Fig. \ref{fig_7}).

 Attempts to explain analytically the observed double-peaks picture on the basis of a classical chain consisting of three spins are provided in the Appendix. Three and five spins are considered to include nearest and next-nearest ferromagnetic DDIs. While the linearized solution for the positions of resonances \cite{HSu55} allows to clarify the appearance of multiple peaks resulting from DDIs between nearest and next-nearest neighbors  and also the relative peaks position with respect to the macrospin case  (Fig. \ref{fig_8}), this model fails  to quantitatively explain the double peaks development shown in Fig. \ref{fig_6}. We therefore adhere to the initial interpretation for the doubled peak stemming from the sizable effective surface anisotropy induced by the nanocube's sides.

For further studies on the system of Fe/Fe$_\mathrm{x}$O$_{\mathrm{y}}$-core-shell-magnetic nanoparticles, experimental information on the oxidation state of the oxide-shell would enable to model the system more appropriately, for  iron-oxide has both ferromagnetic and antiferromagnetic configurations depending on the chemical composition.
Attaining a proper micromagnetic model, the arrays' spectra will result as a  superposition of the single nanocubes' spectra, dominated by the demagnetizing field driven vortex-like configurations and dipole-dipole interaction mediated behavior of the cubes' ensemble.

\section{Acknowledgements}
This work was supported by the German Research Foundation (Nos. SFB 762 and BE 2161/5-1).

\section{Appendix}
We consider a simplified model and compare the resonance's position of a single macrospin with that of a short chain of three macrospins. The considered system's free energy can be written as $F_\Sigma=F_\text{MCA}+F_\text{DDI}+F_\text{ZMN}$, as before with the magnetocrystalline and Zeeman-contribution, however only without the exchange term, since the macrospins are separated  from each other, and also with the DDI term $F_\text{DDI}$, which reads
$
	F_\text{DDI}=\frac{\mu_0}{4\pi}\sum_{i\neq j}\left[\frac{\vec{M}_i\cdot\vec{M}_j-3(\vec{M}_i\cdot\vec{e}_{ij})(\vec{e}_{ij}\cdot\vec{M}_j)}{n_{ij}^3} \right]\ .
$
Here $n_{ij}=r_{ij}/a$ denotes the distance between two spins in units of the edge-length of the cube, i.e. $n_{ij}>1$.

The resonance condition for $N_s$ spins in a linearized solution of the macrospin model is \cite{HSu55}
\begin{align*}
	\left(\frac{\omega}{\gamma} \right)^2=\frac{1}{(N_s M_s)^2\sin^2\Theta}\left[\frac{\partial^2F_\Sigma}{\partial\varphi^2}\frac{\partial^2F_\Sigma}{\partial\theta^2}-\left(\frac{\partial^2F_\Sigma}{\partial\varphi\partial\theta} \right)^2 \right]\ .
\end{align*}

\begin{figure}[htb]
	\centering
		\includegraphics[width=\linewidth]{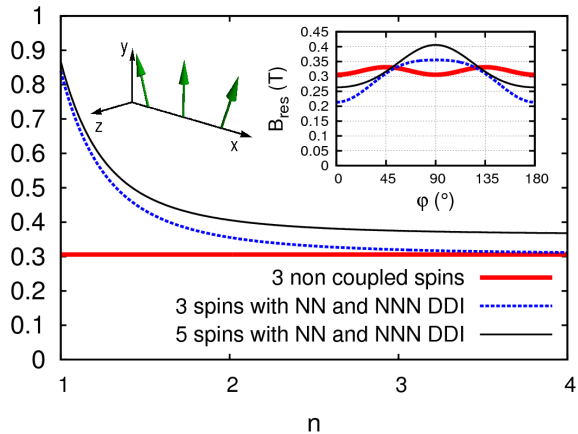}
		\caption{Analytical results for the uncoupled and DDI-mediated resonance position of a spin-chain as a function of the inter-spin distance; the inset shows the angular dependence for different approaches and $\theta=\pi/2$. For the interacting spins nearest and next-nearest neighbours influence is included via DDI. The central spin is aligned with the magnetic field, the following (previous) one is rotated by $\delta_\varphi = \pi/8$ ($\delta_\varphi = -\pi/8$)}
\label{fig_8}
\end{figure}

Aiming to unveil the effect of nearest neighbour (NN) and next-nearest neighbour (NNN) DDI, we show the resulting resonance position in dependence of the considered interactions and also of the spins' distance and quantity.

As the spins are aligned along the x-axis, we can simplify the DDI part of the condition.
In order to describe the vortex structure of the numerical results, a small angular offset between the spins is introduced.

The resulting $\varphi$-dependence for fixed $\theta = 90^\circ$ and $n=\text{distance}/\text{edge-length}=2.0$, can be seen in the inset of Fig. \ref{fig_8}. Here one can observe, that the system's symmetry is reduced from a four-fold to a two-fold one with respect to $\varphi$, due to the interaction between the spins, so that the orientation parallel to the spin-chain becomes less favorable than the orthogonal position, or more precisely, the resonance for the orthogonal orientation requires more energy to be stimulated.
What one can see directly is that the field's resonant strength, even for the uncoupled case is shifted towards higher values for a finite value of $\delta_\varphi$, as the spins are not any more in the direction of an easy magnetocrystalline axis. For the same reason a small difference between the $B_\text{res}$ remains, even for large distances $n$, as the inner energy is increased further for a larger number of spins. In addition, the DDI modifies  the shape of the resonance curve.

Calculating the correction of the DDI to the macrospin-case, one can easily find the resonance-field to be $B_\text{res}=\omega/\gamma-2 K_{c1}/M_s$  for the considered NN- and NNN-cases and for $\delta_\varphi=0$.
\\
\begin{align}
\Delta B_\text{NN}^{N_s=3}
	&= \underbrace{-\frac{\omega}{\gamma}}_{c_1}+\underbrace{\frac{M_s \mu_0}{2 \pi n^3}}_{c_2\propto n^{-3}} +\sqrt{\left(\frac{\omega}{\gamma}\right)^2+\left(\frac{M_s \mu_0}{2\pi n^3}\right)^2} \nonumber
\\
	&= c_1+c_2(n)+\sqrt{c_1^2+c_2^2(n)} \nonumber
\\
\Delta B_\text{NN+NNN}^{N_s=3}
	&= c_1+\frac{17}{16}c_2(n)+\sqrt{c_1^2+\left(\frac{17}{16}c_2(n)\right)^2} \nonumber
\\
\Delta B_\text{NN+NNN}^{N_s=3}
	&= c_1+\frac{21}{16}c_2(n)+\sqrt{c_1^2+\left(\frac{21}{16}c_2(n)\right)^2} \nonumber
\end{align}
For the given values of the constants one observes $c_1=0.321\,\text{T}$, and $c_2=0.352\,\text{T}/n^3$.
\\
Obviously the position of the resonance fields depends crucially  on the choice of the parameters.
Nevertheless generally, the DDI  may deform the resonance line.
The direction of the shift depends on the relative orientation of the chain and the external field (cf. Fig. \ref{fig_8}). 
In the displayed set-up the DDI leads to an increase of the resonance field.
We observe an opposite effect for 
the DDI-mediated coupling of stacked spins in the direction of the magnetic field,
 as inferred  from the angular dependent plot of Fig. \ref{fig_8}.
 Thus,
  one can imagine an interplay  of both of these phenomena   resulting in a multiple peak splitting. 
  An interesting point to emphasize is the comparison of the angular dependent line with Fig. \ref{fig_6}.
   Here, one can distinguish two important aspects: As noted  before, the magnetocrystalline anisotropy increases the systems energy 
   for odd multiples of $45^\circ$, 
   whereas the DDI has an adverse  effect. 
   The combination of both results in the broad spectra for aforementioned odd angular positions.


\end{document}